\documentclass[dvips]{arxstspdf}
\usepackage{graphics,flushend}

\volume{25}
\issue{2}
\pubyear{2010}
\firstpage{170}
\lastpage{171}
\doi{10.1214/10-STS308REJ}
\referstodoi{10.1214/09-STS308}

\begin{document}
\begin{frontmatter}

\vspace*{8pt}
\title{Rejoinder: The Future of Indirect Evidence}
\runtitle{Rejoinder}

\begin{aug}
\author{\fnms{Bradley} \snm{Efron}\ead[label=e1]{brad@stat.stanford.edu}}
\runauthor{B. Efron}

\affiliation{Stanford University}

\address{Bradley Efron is Professor, Department of Statistics, Stanford
    University, Stanford, California 94305, USA.}

\end{aug}



\end{frontmatter}

Our three discussants fit an ``ideal statistican'' profile, combining
deep theoretical understanding with serious scientific interests. The
three essays---which are more than commentaries on my article---reflect
in a telling way their different applied interests: Andrew
Gelman in social sciences, Sander Greenland in epidemiology, and
Robert Kass in neuroscience. Readers who share my bad habit of turning
to the discussions first will be well rewarded here, but of course I
hope you will eventually return to the article itself. There the
emphasis is less on specific applications (though they serve as
examples) and more on the development of statistical inference.

\begin{figure}

\includegraphics{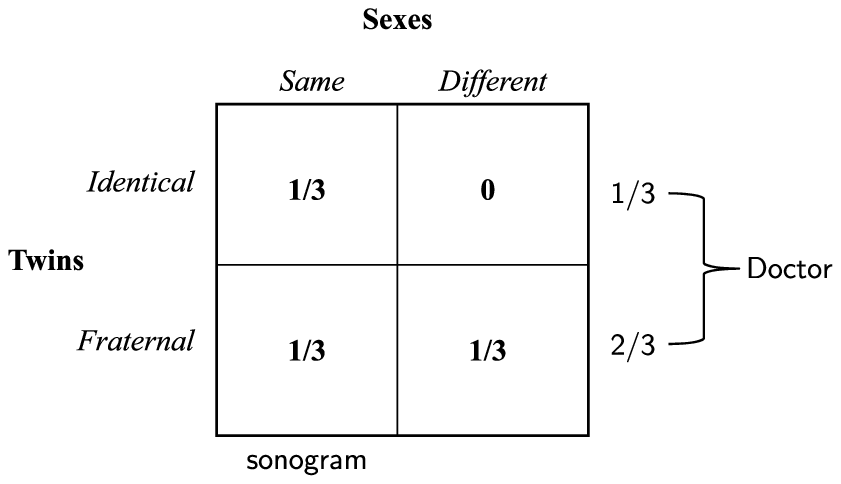}
\vspace*{6pt}
\caption{Probabilities relating to the physicist's twins example of
  Section 3.}
\label{table}
\vspace*{4pt}
\end{figure}

Figure~\ref{table} concerns the physicist's twins example of Section 3. From
the doctor's prior distribution and the fact that sexes differ
randomly for fraternal twins but not for identical ones, we can
calculate probabilities in the four cells of the table. The sonogram
tells the physicist that she is in the left-hand column, where there
are equal odds on identical or fraternal, just as Bayes rule says. In
my terminology, the doctor's indirect evidence is filtered by Bayes
rule to reveal that portion applying directly to the case at hand.
%

There is a leap of faith here, easy enough to make in this case: that
the doctor's information is both relevant and accurate. We would feel
differently if the doctor's evidence turned out to be just three
previous sets of twins, two of which were fraternal. A standard
Bayesian analysis might then start from a $\operatorname{beta}(2,3)$ hyperprior
distribution on the prior probability of \textit{identical}. The
calculation of posterior odds would now be more entertaining than the
actual one in Figure~\ref{table}, but the results less satisfying.

How much respect is due to conclusions that begin with priors, or
hyperpriors, of mathematical convenience? The discussants are divided
here: Gelman, judging from the examples in Chapter 5 of his excellent
book with Carlin, Stern and Rubin, is fully committed; Kass, as a
follower of Jeffries, is mildly agreeable but with strong
reservations; while Greenland seems dismissive (calling objective
Bayes ``please\break don't bother me with the science' Bayes'').

Section 4's empirical Bayes motivation for the\break James--Stein rule
implicitly endorses Gelman's position, except that maximum likelihood
estimation of $M$ and $A$ in (1) finesses the use of a vague
hyperprior for them. The same remark applies to the discussion of
false discovery rates in Section 6. By Section 9, however, my qualms,
along Greenland's lines, become evident: do the estimates $\hat\mu_i$
in Table 2 fully account for selection bias, as they would in a
genuine Bayesian analysis? Kass and I part company here. I believe we
need, and might get, a more complete theory of empirical Bayes
inference while he is satisfied with the present situation, at least
as far as applications go. Gelman is happy with both theory and
applications.

The ground is steadier under our feet for both James--Stein and
Benjamini--Hochberg thanks to\break their frequentist justifications,
Theorems 1 and 2. We do not really need those prior distributions (1)
and (7). The procedures have good consequences guaranteed for
\textit{any} possible prior, which is another way of stating the
frequentist ideal. My ``good work rules'' comment in Section 10 had in
mind the emergence of key ideas such as JS and BH from the frequentist
literature.

Gelman is certainly right: Bayesian statistics has transformed itself
over the past 30 years, riding a hierarchical modeling/MCMC wave
toward a stronger connection with scientific data analysis. This
does not make it an infallible recipe. MCMC methodology has encouraged
the use of mathematically convenient distributions at the hyperprior
level, perhaps a dangerous trend. We could certainly use some new
theory either justifying the recipe or improving upon it.

Maximum likelihood, the crown jewel of classical statistics, is a
theory of direct evidence: the MLE is nearly optimal among nearly
unbiased estimates, while the Fisher information bound tells us how
accurate a direct estimate can be. The most striking lesson of
post-war statistical theory, exemplified by the James--Stein
estimator, is the failure of maximum likelihood estimation in high
dimensions. That failure was the original motivation for this talk and
article, and my (hopefully not futile) call for a more principled
theory of indirect evidence.

``Second-level maximum likelihood'' (using I. J.\ Good's terminology),
as in the empirical Bayes estimation of $M$ and $A$ for the baseball
data, is a tactic for breaking through the MLE dimensional barrier. So
are hierarchical Bayes, random effects models, and regression
techniques. There is no want of methodology here, all of which can be
useful in bringing indirect information to bear, but I find it
difficult to know which methods are appropriate, let alone optimal, in
the analysis of large-scale problems.

The baseball data has outlived several of the players. It has the
sterling virtue of including the ``Truth'' so we can honestly compare
prediction methods. On the downside, nobody cares much about
40-year-old batting averages. We can imagine the same table except
where the proportions refer to cure rates for some horrible disease,
obtained from 18 different experimental drugs. In such a case, pulling
the Clemente of drugs down from 0.400 to 0.294 might seem less
desirable. Relying entirely on direct evidence is an unaffordable
luxury in large-scale data analyses, but indirect evidence can be a
dangerous sword to wield. Some theoretical guidance would be welcome
here, perhaps a theory quantifying the relevance of group data to
individual estimates.

Kass and Gelman rather casually ``dis'' false discovery rates, not on
very good grounds as far as I can see. Fdr methods have done what I
would have thought impossible 15 years ago: displaced Type 1 error
control as the lead technology for large-scale hypothesis testing. Fdr
control is \textit{not} classical significance testing. I consider it
a premonitory example of just the kind of new statistics this article
(and Greenland's essay) hopes for, an amalgam of frequentist and
Bayesian thinking that nicely combines direct and indirect multiple
testing evidence.

I don't mind humility, especially in others, but Kass goes too far in
minimizing his own considerable accomplishments as a scientific
collaborator, and the general role of statistical scientists. Fdr does
\textit{not} ``bless the procedure psychologists were already using.''
The real trick in choosing from a long ordered list of $p$-values is
to know when they stop being interesting. Psychologists (or anyone
else) did not know how to do this trick in 1995 and now they do, thanks
to progress in statistical inference.

Fdr methods can free Kathryn Roeder (as quoted by Kass) from Type 1
error violators' prison. She, and the rest of us, can continue up the
ordered list of $p$-values as far as desired, at each step letting the
local false discovery rate tell her the ever-increasing risk of
misleading her collaborators.

I like Hal Stern's distinction between modelers and nonmodelers,
invoked by Gelman. These days there are three groups to consider,
\begin{eqnarray*}
\mbox{data miners} \ll \mbox{frequentists} \ll \mbox{Bayesians},
\end{eqnarray*}
the inequality signs $\ll$ referring to the amount of probabilistic
modeling. Bayesian modeling is almost always in addition to, rather
than instead of, any frequentist modeling of sampling densities. Data
miners are the atheists of the statistical world, not devoted to
either major philosophy. In fact they often work directly with
algorithms, skipping probabilistic modeling entirely. Good
data-analytic ideas such as boosting and neural networks have come out of
the data-mining/machine learning world (which Rob Kass has at least
one foot in), along with a welcome dose of raw energy. Magical
properties are sometimes attributed to new algorithms---``boosting
methods can never overfit''---before they are digested and
understood in frequentist/Bayesian terms.

Methodology by itself is an ultimately frustrating exercise. A little
statistical philosophy goes a long way but we have had \textit{very}
little in the public forum these days, and I am genuinely grateful to
our editor, David Madigan, for organizing this discussion.
\vadjust{\eject}

\section*{Acknowledgment}

This work was supported in
  part by NIH Grant 8R01 EB002784 and by NSF Grant DMS-08-04324.

\end{document}